\documentclass[
preprint,
tightenlines,
superscriptaddress,
 amsmath,amssymb,
 aps,
]{revtex4-1}

\usepackage{xspace}
\usepackage{graphicx,epsf}
\usepackage{float}
\usepackage{xcolor}
\usepackage{psfrag}
\usepackage{ulem} 

\newcommand{\red}{}                 %

\begin{document}

\title{Thermoelectric current in topological insulator nanowires \red{with impurities}}
\author{Sigurdur I.\ Erlingsson}
\affiliation{School of Science and Engineering, Reykjavik University, 
Menntavegur 1, IS-101 Reykjavik, Iceland}
\author{Jens H.\ Bardarson}
\affiliation{Department of Physics, KTH Royal Institute of Technology, Stockholm, SE-106 91 Sweden}
\author{Andrei Manolescu}
\affiliation{School of Science and Engineering, Reykjavik University, 
Menntavegur 1, IS-101 Reykjavik, Iceland}

\begin{abstract}
In this paper we consider charge current generated by maintaining a temperature difference over a nanowire  at zero voltage bias.
For topological insulator nanowires in a perpendicular magnetic field the current
can change sign as the temperature of one end is increased.  Here we study 
how this thermoelectric current sign reversal depends on  magnetic field and
how impurities affect the size of the thermoelectric current. We consider both scalar and
magnetic impurities and show that their influence on the current are quite similar, although the magnetic impurities seem to be more effective in reducing the effect.  For moderate impurity 
concentration the sign reversal persists.
\end{abstract}

\maketitle

\section{Introduction}
It has been known for quite some time now that the efficiency of thermoelectric
devices can be increased by reducing the system size.  
The size reduction can improve electronic transport properties and also reduce the 
phonon scattering which then leads to increased efficiency \cite{caballero-calero16}.
Interestingly, often the materials that show the best thermoelectric properties on the nanoscale can also exhibit topological insulator properties \cite{penelli14}, although the connection between the two properties is not always straightforward \cite{gooth15}.  Even though few experimental studies exist on thermoelectric properties in topological insulator nanowires (TIN), many studies have reported magnetoresistance oscillations,
both in longitudinal and transversal fields for TINs \cite{Bassler15,Peng:2009jm,Xiu:2011hq,Dufouleur:2013bg,Cho:2015gk,Jauregui:2016cx,Dufouleur17}.

In its simplest form, thermoelectric current is generated when a temperature gradient is maintained across a conducting material.
In the hotter end (reservoir) the particles have higher kinetic energy and thus velocity compared to the colder reservoir.  This leads to a flow of energy from the hot to cold end of the system.  Under normal circumstances this will lead to particles flowing in the same direction as the energy flow.  The charge current can of course be positive or negative depending on the charge of the carriers, i.e.\ whether they are electrons or holes.  Recently, it was shown that in systems showing non-monotonic transmission properties the particle current can change sign as a function of the temperature difference \cite{erlingsson17}.  Sign changes of the thermoelectric current are well know in quantum dots \cite{Beenakker92,Staring93,Dzurak93,Svensson12} when
the chemical potential crosses a resonant state. A sign change of the thermoelectric current can be obtained when the temperature gradient is increased which affects the population of the resonant level in the quantum dot \cite{Svensson13,Sierra14,Stanciu15,Zimbovskaya15}.

For topological insulator nanowires one can expect a reversed, or anomalous, currents
measured in tens of nA \cite{erlingsson17}, well within experimental reach.  Also, since the transport
is over long systems it is much simpler to maintain large temperature difference of tens of Kelvin, compared to the case of quantum dots.
In this paper we extend  previous work on thermoelectric currents in TIN \cite{erlingsson17}, by including the effects of impurities, both scalar and magnetic ones.   The impurities deteriorate the ballistic quantum transport properties, but as long
there are still remnants of the quantized levels, the predicted sign reversal of 
the thermoelectric current remains visible.


\section{Clean nanowires}

When a topological insulator material, e.g.\ BiSe, is formed into a nanowire topological states can appear on its surface.  Such wires in a magnetic field have recently been studied extensively both theoretically~\cite{Bardarson:2010jl,Zhang10:206601,Zhang:2012ci,Ilan:2015ei,Xypakis:2017kw} and experimentally~\cite{Peng:2009jm,Xiu:2011hq,Dufouleur:2013bg,Cho:2015gk,Jauregui:2016cx,Dufouleur17,Arango:2016dd}.
When the nanowires are of circular cross-section the electrons move on a cylindrical surface with radius $R$.
The surface states of the topological insulator are Dirac fermions, described by the Hamiltonian~\cite{Bardarson:2010jl,Zhang10:206601,Bardarson:2013cn}
\begin{equation}
H_\mathrm{TI}=-i\hbar v_F \left[
\sigma_z \left( \partial_z+i\frac{eB}{\hbar}R \sin \varphi \right) +\sigma_y \frac{1}{R}\partial_\varphi
\right ] ,
\label{eq:HamiltonianTI}
\end{equation}
where $v_F$ is the Fermi velocity, and the spinors satisfy anti-periodic boundary conditions $\hat{\psi}(\varphi)=-\hat{\psi}(\varphi+2\pi)$, due to a Berry phase \cite{Bardarson:2010jl,Zhang10:206601}.  Here we chose the coordinate system such that magnetic field is along
the $x$-axis, ${\bf B}=(B,0,0)$, the vector potential being ${\bf A}=(0,0,By)=(0,0,BR\sin\varphi)$.  
For $B=0$ the angular part of the Hamiltonian has eigenfunctions
$e^{i \varphi n}/\sqrt{2\pi}$ where $n$ are half-integers to fulfill the boundary condition.
It is convenient to diagonalize Eq.\ (\ref{eq:HamiltonianTI}) in the angular basis, which are exact eigenstates when 
$B=0$.

\begin{figure}{ } 
\begin{center}
\includegraphics[angle=-90,width=0.48\textwidth]{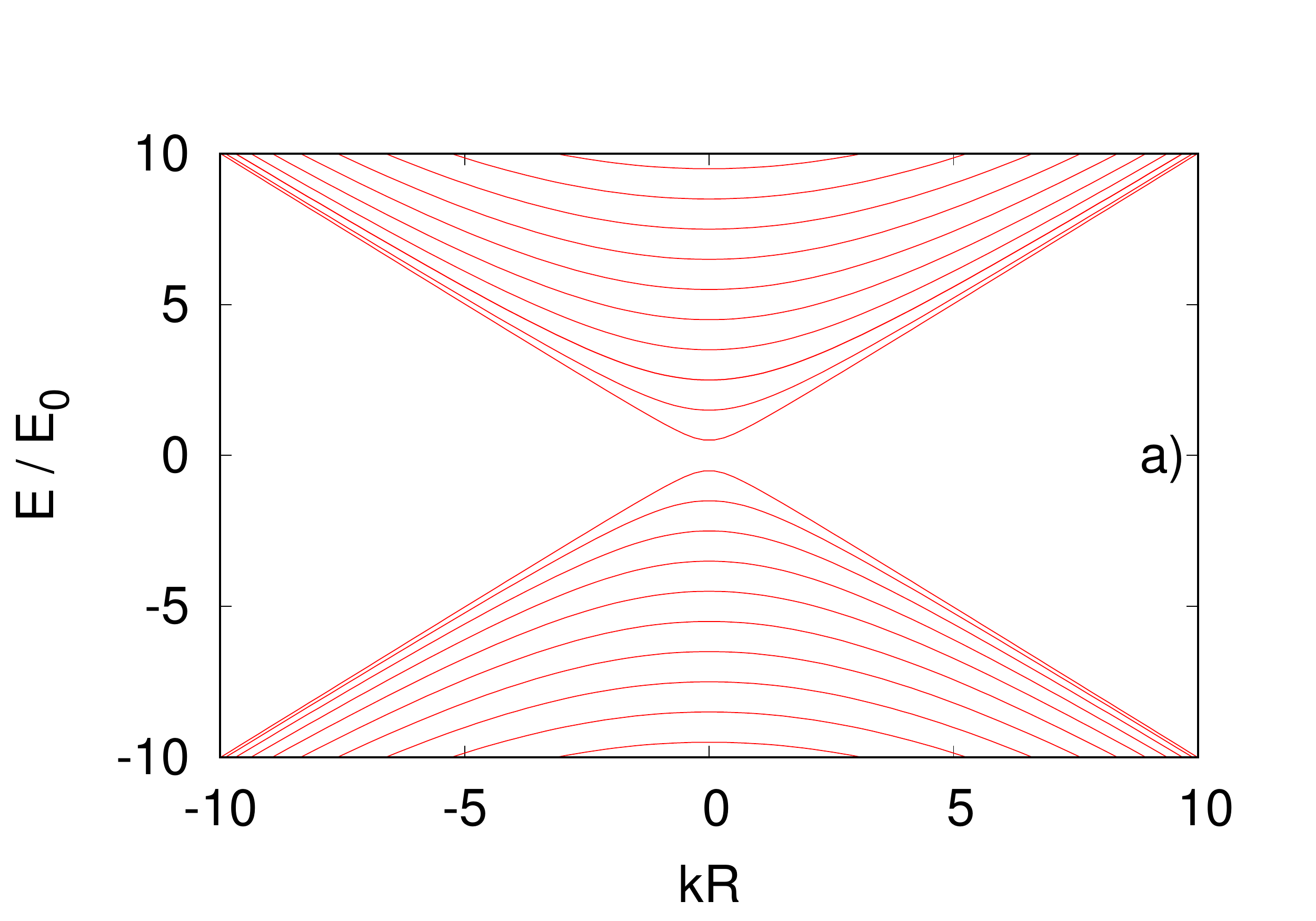}
\includegraphics[angle=-90,width=0.48\textwidth]{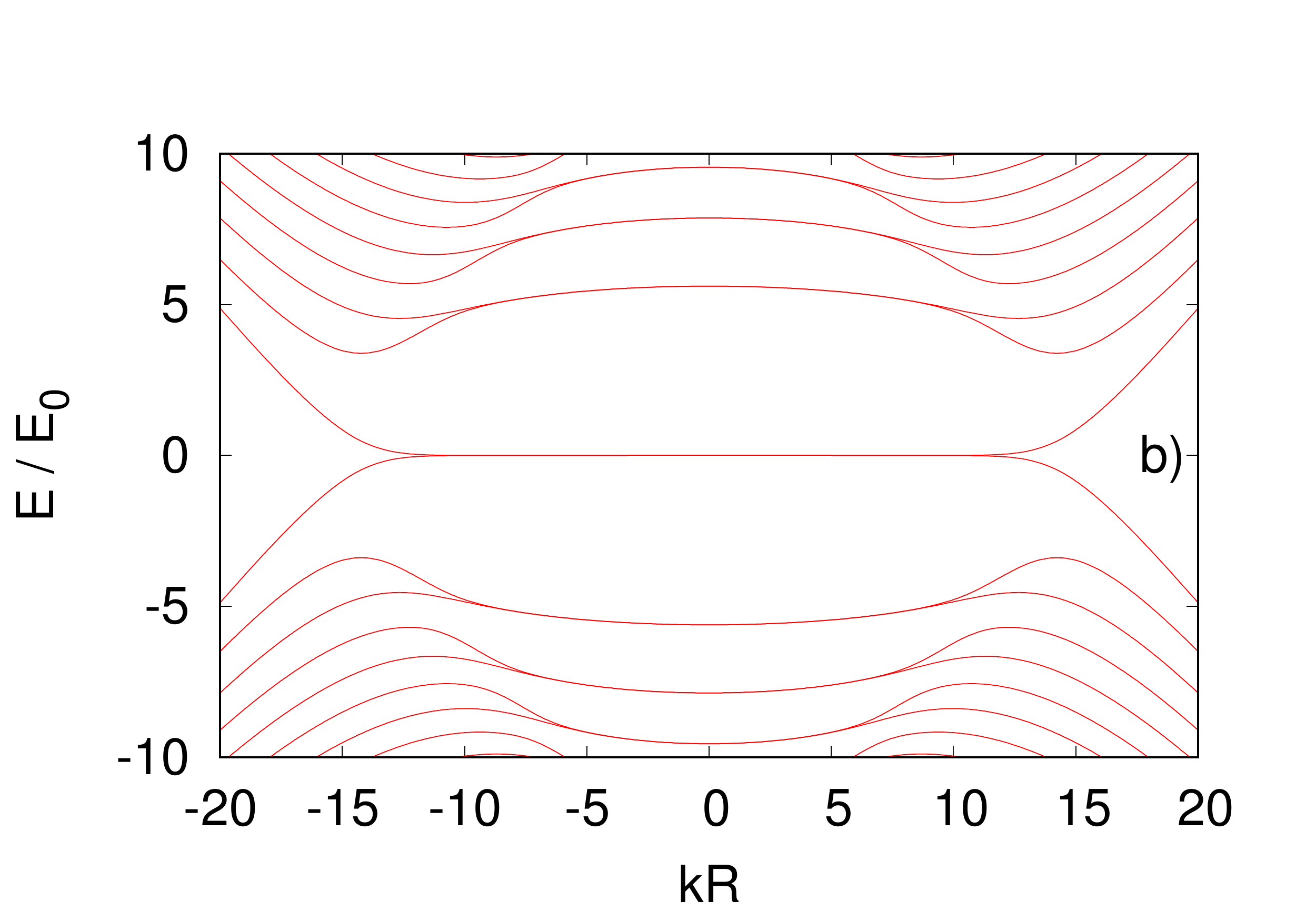}
\vspace{-2mm}
\end{center}
\caption{The energy spectrum for a) $B=0$ and b) $B=4.0$\,T.  Note that the system is gapped at $B=0$, but not at $B=4.0$\,T.  Here we used  $v_F=10^5$\,m/s, $R=50$\,nm for the current calculations, which gives $E_0=\frac{\hbar v_F}{R} \approx 1.3$\,meV.}
\label{fig:TInanowire}
\end{figure}

An example of the energy spectrum is shown in Fig.~\ref{fig:TInanowire} for a) $B=0$ and b) for $B=4.0$\,T.  
\red{The model parameters are comparable to experimental values \cite{Dufouleur17}.
For zero magnetic field the energy spectrum has a gap at $k=0$, 
resulting from the anti-periodic boundary conditions \cite{Bardarson:2010jl,Zhang10:206601}. }
For the non-zero magnetic field case precursors of Landau levels around $k=0$ are seen, both at negative and positive energy. The local minima away from $k=0$ are precursors of snaking states.
Such sates have been studies for quadratic dispersion (Schr\"odinger) where the Lorentz force always bends the electron trajectory towards
the line of vanishing radial component of the magnetic field. \cite{Tserkovnyak06,Ferrari09,Manolescu13,Chang16}.
\red{In fact this is a classical effect known in the two-dimensional electron gas in inhomogeneous magnetic
fields with sign change \cite{Muller92,Ibrahim95,Ye95,Zwerschke99}.  For Dirac electrons it has been reported
in graphene p-n junctions in a homogeneous magnetic field, since in this case the charge carriers change sign \cite{Rickhaus15}.}

In order to calculate the current in multi-channels one dimensional systems one needs
to calculate the product of the velocity $v_n(E)$ and density of states $\rho_n(E)$ of a given mode $n$ at energy $E$ \cite{nazarov05}.  This product  is a constant $v_n(E) \rho_n(E)=\frac{1}{h}$, irrespective of the form of $\varepsilon_{n}(k)$, which leads to the well known conductance quantum $\frac{e^2}{h}$.  For infinitely long, ballistic systems all channels are perfectly transmitted $T_n=1$, so one can simply count the number of propagating mode to get the conductance.
If the curvature of the dispersion is negative (here we consider positive energy states) at $k=0$, then the mode can contribute {\it twice} to the conductance since there are two values of $k$ that fulfill $\varepsilon_{n}(k)=E$ that have the same sign of $v_n(E)$, see Fig.\ \ref{fig:TInanowire}b).
The transmission, which in this case is simply the number of propagating modes, can jump up by two units and then again down by one unit, as a function of energy.
%
As was pointed out recently, the presence of such non-monotonic behavior in the transmission function $T(E)$ can give rise to {\it anomalous}
thermoelectric current\cite{erlingsson17}.

\begin{figure}
\begin{center}
\includegraphics[angle=-90,width=0.48\textwidth]{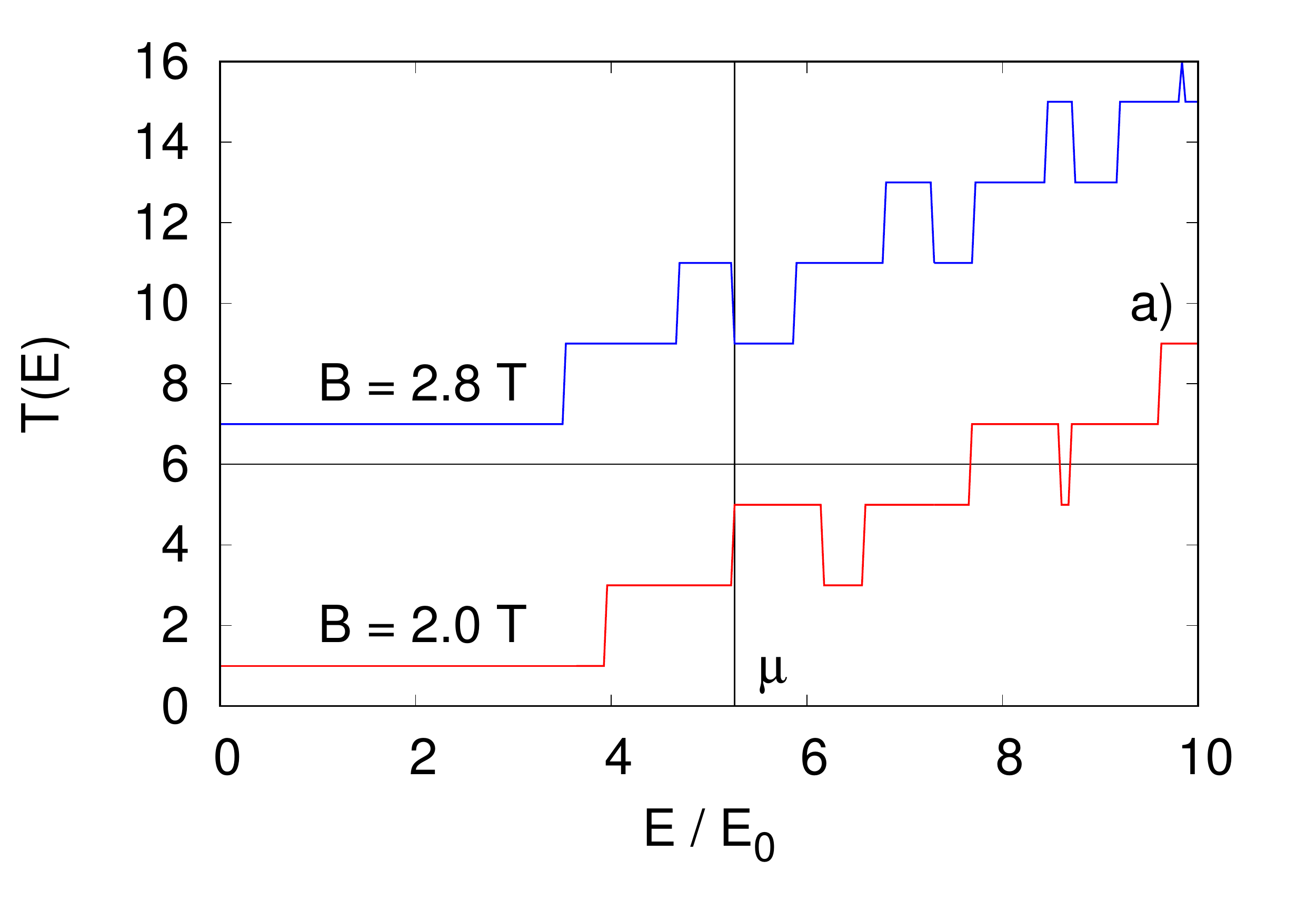}
\includegraphics[angle=-90,width=0.48\textwidth]{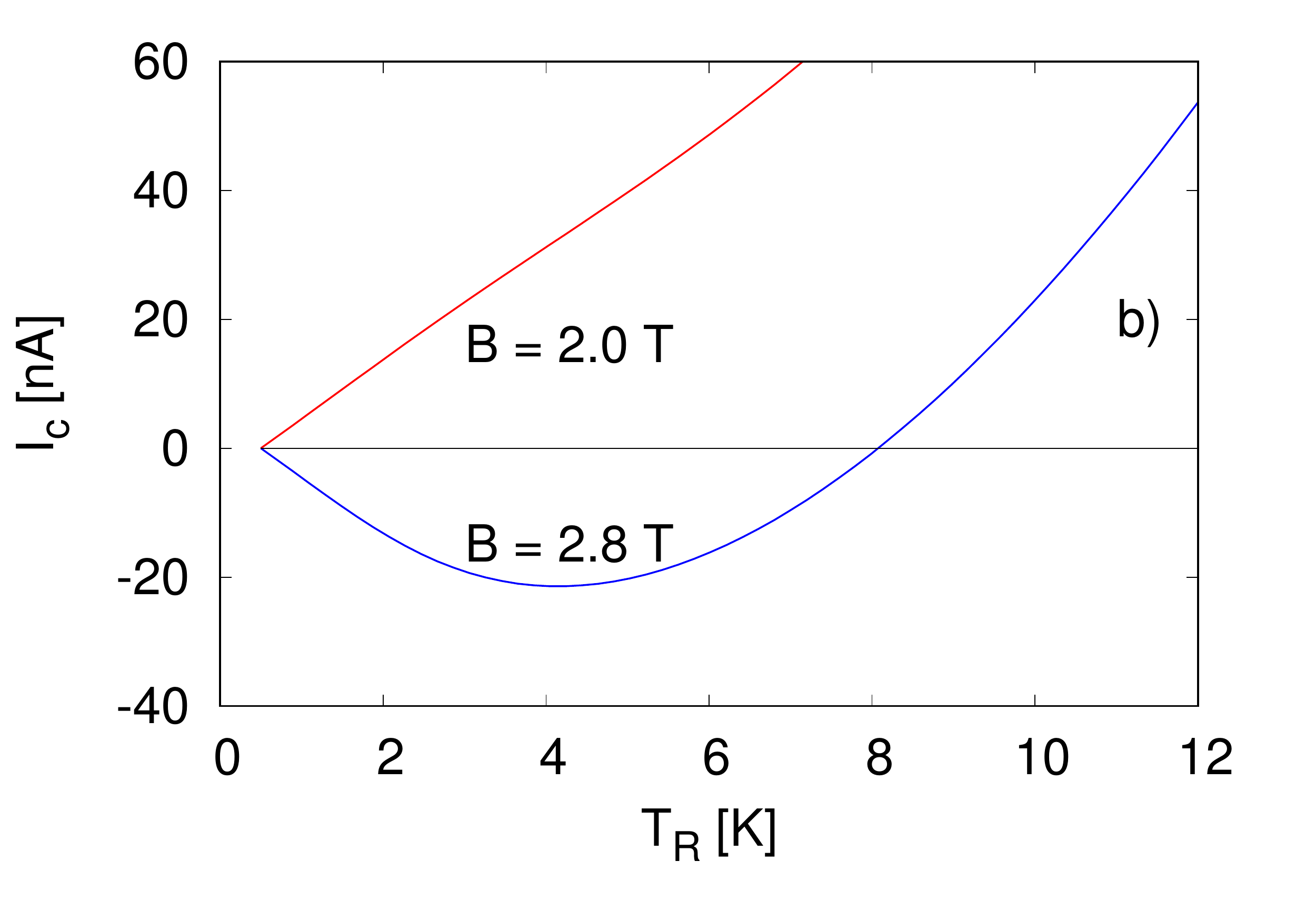}
\vspace{-2mm}
\end{center}
\caption{The transmission function is shown in a) and the thermoelectric current b), for two different magnetic fields.  In a) the transmission function $T(E)$ for $B=2.8$\,T is offset by 6 for clarity.  Here we used  $v_F=10^5$\,m/s, $R=50$\,nm for the current calculations, which gives $E_0=\frac{\hbar v_F}{R} \approx 1.3$\,meV.}
\label{fig:TEIc_Temp_Bfield}
\end{figure}

In order clarify the origin of the sign reversal of the thermoelectric current, and 
how its affected by magnetic field, we will briefly outline how the current is calculated using the Landauer formula.
The charge current $I_c$ is given by
\begin{equation}
I_{\rm c}=\frac{e}{h}\int T(E) \left[f_{\rm R}(E)-f_{\rm L}(E)\right] dE\ .
\label{eq:Ic}
\end{equation}
Here $f_{\rm L/R}(E)$ are the Fermi functions for the left/right reservoir
with chemical potentials $\mu_{\rm L/R}$ and temperatures $T_{\rm L/R}$.  Here we will consider $\mu_L=\mu_R=\mu$.
If the transmission function $T(E)$ increases
with energy over the integration interval (and the chemical potential is situated somewhere in the interval) the thermoelectric current
is positive.  This is the normal situation.  
An anomalous negative current can instead
occur if the transmission function decreases with energy. The curve for $B=2.0$\,T  in
Fig.~\ref{fig:TEIc_Temp_Bfield}a) shows the normal situation where the chemical potential is positioned at an {\it upward} step at $\mu=6.8$\,meV.
The vertical line indicates the position of $\mu$.  The resulting charge current is shown in Fig.~\ref{fig:TEIc_Temp_Bfield}b) where the normal situation is evident for $B=2.0$\,T.
If the magnetic field is increased to $B=2.8$\, the energy spectrum changes (not shown) and so will the transmission function $T(E)$.  Now a downward step occurs at $\mu$ which leads to an anomalous current, as can be seen in  Fig.~\ref{fig:TEIc_Temp_Bfield}b).
Note that the current can change sign by either varying the temperature of the right reservoir or the magnetic field. 
The anomalous current can be in the range of tens of nA, which is well within experimental reach.
The anomalous current can be in the range of tens of nA, which is well within experimental reach.

\section{Impurity modelling}

The anomalous current introduced above relies on non-monotonic steps
in the transmission function. For ballistic nanowires the steps are sharp, but
in the presence of impurities the steps will get distorted.
In order to simulate transport in a realistic nanowires, we will assume short range impurities.  These are described by
\begin{equation}
 V_\mathrm{imp}(z,\varphi)=\sum_i W\delta(z-z_i)\delta(\varphi-\varphi_i)\ ,
 \label{eq:Vimp}
\end{equation}
where $W$ is the impurity strength.  Due to Fermion doubling the Hamiltonian in Eq.\ (\ref{eq:HamiltonianTI}) can not be directly discretized \cite{stacey82}.  However, adding a fictitious quadratic term 
\begin{equation}
 H_\lambda=-\frac{v_\mathrm{F} \hbar}{R} \lambda^2 \sigma_x (R^2 \partial_z^2) 
\end{equation}
solves the Fermion doubling issue \cite{masumhabib16}.  To fix the value of $\lambda$, we will first look at the {\it longitudinal} part of Eq.\ (\ref{eq:HamiltonianTI}) in the absence of magnetic field
\begin{equation}
H_\mathrm{TI,z}=-i\hbar v_F 
\sigma_z \partial_z  -\frac{v_\mathrm{F} \hbar}{R} \lambda^2 \sigma_x (R^2 \partial_z^2) .
\label{eq:HamiltonianTI_z}
\end{equation}
If this Hamiltonian is discretized on a lattice with a lattice parameter $a$ the spectrum will be
\begin{eqnarray}
 \varepsilon_{\pm}(k)&=& \pm \frac{\hbar v_\mathrm{F}}{R} \sqrt{\frac{R^2}{a^2}\sin^2 \left ( ka \right )+(2\lambda)^4\frac{R^4}{a^4}\sin^4 \left ( \frac{ka}{2} \right )}
 \label{eq:ek}
\end{eqnarray}
where $ka \in [-\pi,\pi]$.  The value of $\lambda$ can be set by the condition that the Taylor expansion of $(\varepsilon_\pm(k))^2$ contains no quartic term, which maximizes the region showing linear dispersion.  This condition is fulfilled when
\begin{equation}
 \lambda =\frac{1}{3^{1/4}}\sqrt{\frac{a}{R}} \ .
 \label{eq:lambda}
\end{equation}
\red{For zero magnetic field we choose the lattice parameter $a=0.02R$, which ensures that the first ten
states calculated via the lattice model with the $\lambda^2$ term 
deviate by less than 1\% from those obtained with the continuum model (Fig.\ \ref{fig:TInanowire}a).  For a non-zero magnetic
field we use $a=0.01R$, because more states contribute to the flat bands at $E=0$.}
At this point we are free to use standard discretization schemes and the transmission function in the case when impurities
are included is obtained using the recursive Green's function method\cite{Ferry97}.

Experiments on normal (not topological) nanowires show a conductance 
that can be a complicated, but reproducible trace for a given nanowire.
This means that the measurement can be repeated
on the same nanowire and it will give the same conductance
trace as long as the sample is kept under unchanged conditions.  But a different nanowire would show a different, but reproducible, conductance trace \cite{Wu13:4080}.
This motivates us to consider a fixed impurity configuration, i.e., no ensemble average. 

\begin{figure} 
\begin{center}
\includegraphics[angle=-90,width=0.48\textwidth]{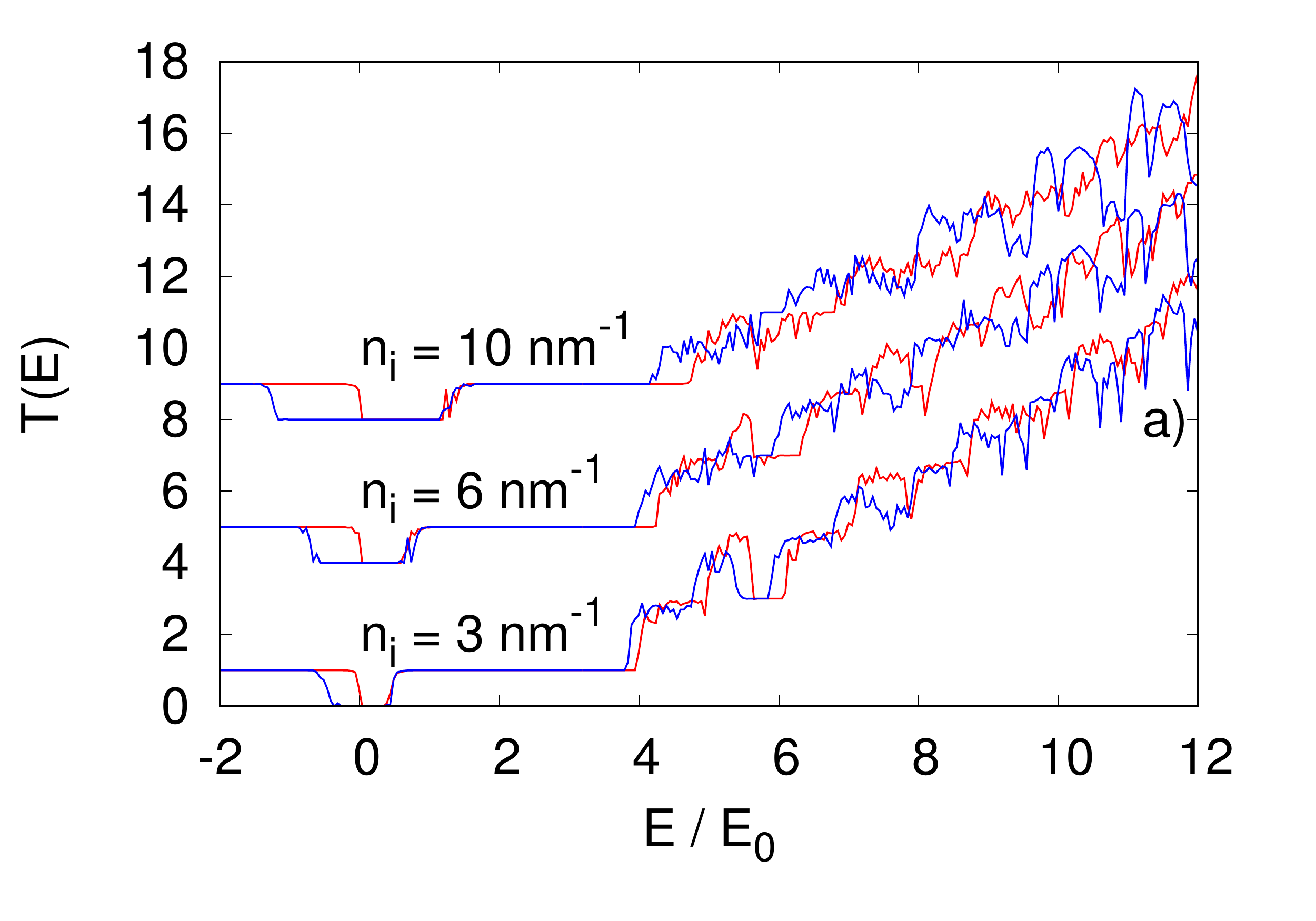}
\includegraphics[angle=-90,width=0.48\textwidth]{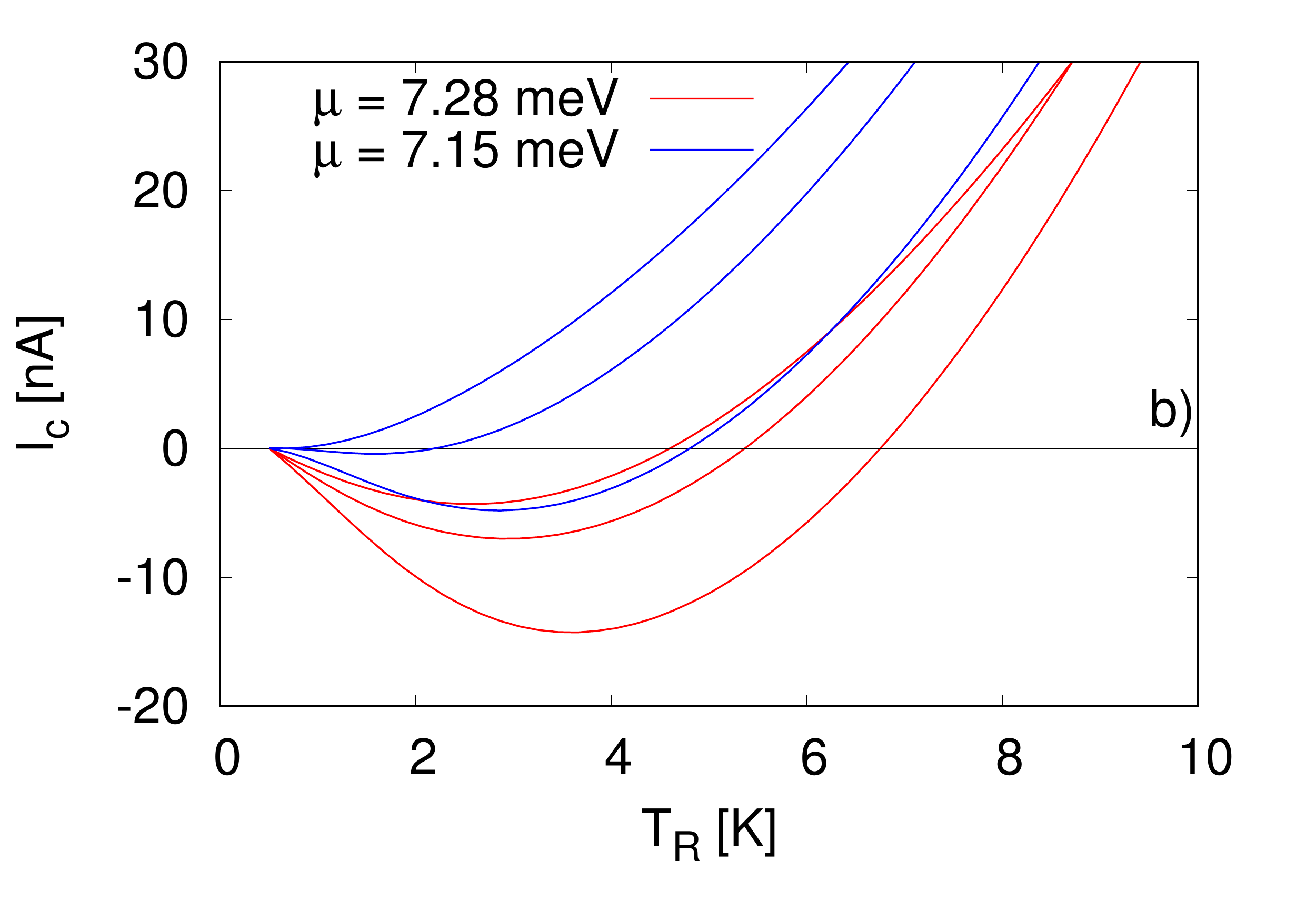}
\end{center}
\vspace{-1mm}
\caption{Transmission function a) and thermoelectric current b) calculated in the presence of impurities at $B=4.0$\,T.
The nanowire length is $L=1000$\,nm and the impurity densites are $n_i=3.0$\,nm$^{-1}$, 6.0\,nm$^{-1}$, and 12\,nm$^{-1}$.
The red curves are for scalar impurities, with chemical potential $\mu=7.28$\,meV, and 
the blue curves are for magnetic impurities, with $\mu=7.15$\,meV.}
\label{fig:current_imp}
\end{figure}

In Fig.~\ref{fig:current_imp} we show the transmission functions and the thermoelectric currents for 
magnetic field  $B=4.0$\,T, 
for a nanowire of length $L=1000$\,nm.  The disorder strength is set to $W=4.8 \frac{\hbar v_F}{R}$ and the density of impurities is varied $n_i=3.0$\,nm$^{-1}$, 6.0\,nm$^{-1}$ and 12\,nm$^{-1}$.
For comparison we consider two types of impurities: scalar impurities described by Eq.\ (\ref{eq:Vimp}) [red traces], and magnetic impurities described by $V_\mathrm{imp} \sigma_x$ [blue traces].
When the transmission function in Fig.~\ref{fig:current_imp}a) in the presence of impurities is studied a definite trend towards
reduced non-monotonic intervals is visible as the density of impurities is increased from $n_i=3.0$ to 6.0\,nm$^{-1}$ and 12\,nm$^{-1}$.  This applies both to scalar [red] and magnetic impurities [blue],
even though the magnetic impurities seem to cause a quicker reduction in the transmission peaks.  The impurities, both scalar and magnetic, open up a gap around $E=0$.  This is due to scattering between counter propagating states on the same side of the nanowire \cite{Xypakis:2017kw}.

When looking at the calculated charge current current in Fig.~\ref{fig:current_imp}b) the difference between the scalar and magnetic impurities becomes more clear.  In both cases the strength and density of impurities is the same but in the magnetic case the impurities are substantially more effective in reducing the anomalous current.  Note that due to different impurity configuration between the magnetic and scalar cases we adjusted the chemical potential to $\mu=7.15$\,meV to maximize the anomalous current. The values of $W_\mathrm{imp}$ and $n_i$  used here, were chosen such that we could observe an evolution in Fig.~\ref{fig:current_imp}a) from resolving the quantized steps to not seeing any.  For experiments, this would mean that samples that show some indication of quantized conductance steps should suffice to observe the anomalous current.

\red{In our calculations we neglected the 
Coulomb interactions between electrons, which, 
in the nonlinear regime of transport, may alter the current, at least in non-topological 
materials \cite{Sanchez16,Sierra16,Torfason13}. To our knowledge, the present experimental data in TI nanowires can
be explained without considering the Coulomb interaction. But, nevertheless, this issue can be an open question
for future research.
}

\section{Conclusion}
We studied reversal of thermoelectric current in topological insulator nanowires and how it evolves with changing magnetic field.  Using lattice models we simulated realistic nanowires with both scalar and magnetic impurities.  Even though both scalar and magnetic impurities reduce the size of the anomalous current we expect that in quasiballistic samples the effect should be observable.  Interestingly the magnetic impurities seem to be more effective than the scalar ones when it comes to reducing the anomalous thermoelectric current.
\red{For hollow nanowires described by the Schrödinger equation
the backscattering is the same for magnetic and scalar impurities,
in the absence of spin-orbit interaction. This is in contrast to the
TI nanowires studies here which are more susceptible to scattering by
magnetic impurities compared to scalar ones, due to spin-momentum locking
of the surface states \cite{Ilan:2015ei}. }

\acknowledgments
{This work was supported by: RU Fund 815051 TVD and ERC Starting Grant 679722. 
}

\bibliography{thermo_beiler}

\end{document}